# Reasons, Challenges and Some Tools for Doing Reproducible Research in Transportation Research

Zuduo Zheng

School of Civil Engineering, the University of Queensland, St Lucia, Queensland, Australia 4072

**Abstract**: This paper introduces reproducible research, and explains its importance, benefits and challenges. Some important tools for conducting reproducible research in Transportation Research are also introduced. Moreover, the source code for generating this paper has been designed in a way so that it can be used as a template for researchers to write their future journal papers as dynamic and reproducible documents.

## Introduction

How to validate a scientific finding is a challenge as old as human's scientific activity itself. Replication is generally regarded as the gold standard of validating a scientific study. Unfortunately, replicating a study is often very difficult if not impossible at all due to various reasons. On the other hand, research validation is becoming important and urgent more than ever. Just think about the huge amount of data we are collecting, the high complexity of the algorithms we are developing, the unprecedented computational power we are using on a daily base, the fast-paced (it's getting even faster!) publication industry, just to name a few.

Reproducible research (RR) is a compromise of replication, rather than a replacement of replication. Instead of validating a study (this is the main mission of replication), reproducible research aims to validate a study's data analysis. Simply speaking, reproducible research is an extra (but very important) effort from the authors of a publication to share their data, codes, and instructions on how to piece them together for the purpose of enabling a third party to easily and conveniently obtain the identical results reported in their paper. The definition of RR in the Wikipedia entry is:

> **"The term reproducible research refers to the idea that the ultimate product of academic research is the paper along with the laboratory notebooks and full computational environment used to produce the results in the paper such as the code, data, etc. that can be used to reproduce the results and create new work based on the research."**

(accessed on the 9th of January 2019; link: https://en.wikipedia.org/wiki/Reproducibility#Reproducible_research)

As implied in the definition above, the primary target of RR is computational analysis, which is deeply rooted in numerous disciplines thanks to the rapid advancement of computing technologies and the indispensable role of computing in the modern science and engineering, including Transportation Research. In Transportation Research, computational analysis is ubiquitous and woven into our daily activities, either as a researcher, a practitioner, or as a student. The penetration of the computational analysis in Transportation Research is spreading out to every corner, e.g., statistical and econometric models for road safety analysis and travel behaviour analysis; mathematical and simulation models for traffic flow theories, traffic operation and control; optimization methods for transportation network analysis, just to name a few. By developing and relying more and more on computational algorithms we hope that the computational part of our work can be readily implemented by ourselves, easily disseminated to and precisely reproduced by other researchers, and conveniently modified, extended or enhanced in future studies. Ironically, what is happening could not be further away from the original goal, and many computational analyses become a black box (knowingly or unknowingly to the authors) as the algorithms used become ever larger and more complex, more data-thirsty, and more intelligent, which makes even precisely documenting an algorithm itself often challenging. The gap between what was actually done and what is described in the published paper is often so unfillable that reproducing the analysis upon which main findings are based is extremely difficult. Sometimes seemingly-trivial details that are often deemed (either by authors or by reviewers) as not worth reporting in a published paper can make repeating the analysis unattainable or immensely time-consuming (thus, discouraging any reproducing effort) (Boettiger 2015), e.g., a threshold used for filtering outliers in the data, initial values and termination criteria used for a numerical optimization algorithm, even instruction on how to install the software used in the paper. Even worse, it is not rare that the actual code or software "mashup" for producing the final analysis may be lost or unrecoverable (Mesirov 2010). More and more researchers have realized that letting the status quo unchanged would jeopardize the credibility of computational analysis.

**"Computation-based science publication is currently a doubtful enterprise because there is not enough support for identifying and rooting out sources of error in computational work"**

— David L. Donoho (Donoho 2010)

Although Donoho was talking about computational science, it also applies to many research activities in Transportation Research. To address this paradox in computational analysis, RR is gaining unprecedented momentum in several disciplines, as discussed later. Benefits of RR are multifaceted, and well documented in the literature (Donoho 2010; Peng 2011; Gandrud 2016), as outlined below.

Main benefits of RR to you as an author include:

- Greater impact of your research. Making your research reproducible can increase your reputation as a researcher, and help attract more citations of your work.
- Improved work and work habits, and improved teamwork. Making your research reproducible can make you more efficient and help you minimize the chance of error.

Main benefits of RR to your readers:

- Improved quality and trustworthiness of the research they are reading.
- Easy reproducibility of the analysis.
- Easy extension of the analysis for further investigation on a related topic. This is also true to the author(s).
- Improved productivity.

Main benefits of RR to the society include:

- Stewardship of public goods. RR has an irreplaceable role in curbing and preventing academic misconducts, frauds, and scandals.
- Increased public access to public goods.
- Better quality of the education system. RR can also be employed in our lectures and teaching to better educate students and the next generation of researchers by encouraging them to interact with the paper they are reading by repeating the entire or part of the analysis to reproduce the result, rather than being a passive, often intimidated consumer. Such practice itself can increase their awareness of RR, and make RR the norm of their daily practice after they graduate.

An academic paper based on computational analysis typically consists of two main components: the data analysis or modelling results and narratives for explaining, advertising and promoting these results. Some readers may disagree with the use of "advertising" and "promoting" to describe an academic paper. As a matter of fact, any journal paper is essentially just for advertising a research work the authors have completed, no more and no less. As Buckheit and Donoho once said,

**"An article about computational science in a scientific publication is not the scholarship itself, it is merely advertising of the scholarship. The actual scholarship is the complete software development environment and the complete set of instructions which generated the figures."**

— John Claerbout, paraphrased in Buckheit and Donoho (1995), sometimes referred to as Claerbout's Principle (De Leeuw 2001).

Correspondingly, the author(s) of a journal paper has two main tasks: writing the codes to implement the data analysis or modelling in order to generate the results for being included in the paper, and writing the narratives to explain what(s) (e.g., what is the objective, what are the research questions), how(s) (e.g., how is each research question answered), and why(s) (e.g., why is a particular method chosen), which is very much beyond the traditional way of commenting the codes as commonly practiced in computer science related fields. To help the author(s) to efficiently (or even effortlessly) accomplish these two tasks, an ideal tool should have the functionality of allowing the authors to write a paper as they usually do, but replace the analysis/modelling results (including byproducts of the analysis like figures, tables) with the source codes. During the writing, the authors do not need to worry about any formatting issues, and only concentrate on the content itself (as a writer should do). At the end of the writing, the narratives and the source codes can be mixed together by the tool to generate the paper in a user-chosen format (e.g., PDF, MS Word, html, etc.) with a ready-for-submission quality. There are many notable endeavours for developing such a powerful tool, and some are closer to reaching the goal than others, as discussed later.

We are at a crossroads where how academic work should be conducted, documented and disseminated, and RR is an unstoppable and inevitable future. While such belief is not new and has been gaining a massive following in many disciplines (e.g., Biostatistics, signal processing, statistical analysis, etc.), it is by and large unnoticed in the Transportation Research community. This paper aims to introduce basic elements of RR to researchers in Transportation Research and facilitate this transition in the Transportation Research community. Towards this end, remainder of this paper is organized as follows. Section 2 reviews the history of RR and RR practice in other disciplines. Section 3 introduces some tools (e.g., RStudio, R Markdown) freely available for easily conducting RR. Section 4 presents an example of using these tools to convert one of my previous publications into a RR form. Section 5 discusses opportunities and challenges of doing RR. Finally, Section 6 concludes this article by summarizing the main points.

Note that this article was written entirely in a reproducible way using **Rstudio** and **RMarkdown** from head to toe. **The source code for generating the exact document you are reading was submitted to the journal as supplementary material for the review purpose**, and can be downloaded from GitHub: xxx (the link will be added once the paper is accepted for publication) or from the webpage: http://www.connectedandautonomoustransport.com/xxx (the link will be updated once the paper is accepted for publication).

## A brief history and recent development of reproducible research

It is difficult to trace back the origin of RR. It is reasonable to assume that the concern on a scientific work's reproducibility started as early as the origin of science, while Robert Boyle (a natural philosopher and a pioneer of modern experimental scientific methods) started the notion of reproducibility as a scientific standard in the 1660s (LeVeque, Mitchell, and Stodden 2012). To researchers in some disciplines, making their work reproducible is the minimum expectation, and the necessity of reproducibility is deeply rooted with a long history, e.g., for mathematicians, they need to prove the correctness of any theorem they have discovered. The purpose of "the proof" of a mathematical theorem is to ensure the reproducibility of this piece of work. Experiments by physicists and chemists need to be independently reproduced by other researchers. In contrast, researchers in fields related to computational science did not pay much attention to reproducibility until early 1990s.

Generally, people believe the term "reproducible research" was coined by Jon Claerbout, a geophysicist and Professor at Stanford University (Fomel and Claerbout, 2009). Quickly, it gained lots of attention from the computational science community. For instance, one paper by Roger Peng on RR was published in Science in 2011 (Peng 2011). Several books, special issues, papers for advocating RR in several disciplines have been published ever since. Guidelines on RR have been proposed by many researchers with different focuses. Moreover, some tools that are open source and freely available have been specifically developed for facilitating RR. However, by and large, to date the Transportation Research community has not paid much attention to this RR movement.

Although to many researchers in Transportation Research RR is perhaps still a new concept, disciplines like biology, biostatistics, signal processing (Peng 2009) have been making great strides. Some notable efforts are described below.

As perhaps the first tool specifically designed for the purpose of reproducing previous analysis, and inspired by Jon Claerbout, Buckheit and Donoho (1995) developed Wavelab, a library of wavelet-packet analysis, cosine-packet analysis and matching pursuit. Particularly, using Wavelab users can reproduce all the figures in their published wavelet articles. Fadili et al. (2010) developed MCALab for conducting RR in sparse-representation-based signal and image decomposition and inpainting. For analyzing genomic data, Mesirov (2010) created a RR system (GenePattern) with two components: a Reproducible Research Environment for doing the computational work and a Reproducible Research Publisher for documenting and publishing the analysis. Goecks et al. (Goecks, Nekrutenko, and Taylor 2010) developed a tool called Galaxy for supporting accessible, reproducible, and transparent computational research in the life sciences. Chaumont et al. (De Chaumont et al. 2012) developed an open platform (called Icy) for RR in bioimage informatics. Peng et al. (2006) outlined a standard for reproducibility and evaluated the reproducibility of current epidemiologic research. Donoho et al. (2009) discussed RR in computational harmonic analysis. Vandewalle et al. (2009) surveyed RR in signal processing, and introduced a reproducible research repository where one can publish one's code, rate a given paper's reproducibility.

Besides these domain specific tools for RR, significant progresses on developing general RR tools have occurred, too. A software framework based on the concept of compendium (a container for one or more dynamic documents and the different elements needed when processing them) was proposed by (Gentleman and Lang 2007). Howe (2012) propose a model of RR that performs experiments within a virtual machine hosted by a cloud provider (this approach causes many new issues such as cost, security, reuse, limitations to interactivity, etc.). An open source tool Docker has been designed for computational reproducibility, and is gaining popularity (Boettiger 2015). A package for using R on Docker called Rocker is available (Eddelbuettel 2015), too.

Meanwhile, some researchers also investigated legal issues that may hinder RR's acceptance. For instance, Stodden (2009) proposed a legal framework, the Reproducible Research Standard, to encourage scientific research by rescinding the aspects of copyright that prevent scientists from sharing important research information, and thus enables reproducibility. How RR should be cited is also discussed in the literature (Gandrud 2016).

Overall, despite many notable endeavours and significant progresses for turning RR into a norm, most of them did not attract mainstream attentions because they are not easy to use, and learning these tools itself poses a serious challenge. Recently, an array of RR-oriented packages have been developed using the free and open-source language **R** (https://www.r-project.org/); among them the most notable ones are **knitr**, **R Markdown**, and **bookdown**. Users of **RStudio** (https://www.rstudio.com/; a free and open-source integrated development environment for **R**) can produce a dynamic document using the simple syntax of **R Markdown** (more discussion later), and then weave it into a final output of various types (e.g., PDF, html, or a word document) with a single click. Users can focus on doing the data analysis and writing the content as academic writings should do without the need of worrying technicalities related to dynamic document/literate programming (e.g., tangling, weaving; more discussion later) because these issues are silently taken care of by **knitr** or its newly developed cousin **bookdown** (a word play of **Markdown**, which itself is a word play of **Markup**; developed by Yihui Xie, the developer of **knitr**). Catalyzed by these powerful, user-friendly and freely-available tools, RR is at the verge of becoming a new norm of how researchers should write any computational analysis related documents and how publisher should publish such work. The primary goal of this paper is to increase the awareness of researchers in Transportation Research on the most recent development of RR in other disciplines and introduce to them some easy-to-use tools to make their journey of transitioning to RR smoother.

## How to conduct reproducible research

The idea of RR is akin to the idea of literate programming proposed in 1984 by computer scientist and mathematician Donald E. Knuth for a better documentation of programs by considering programs to be *works of literature* (Knuth 1984). This is where the name *literate programming* comes from. A literate program is a document that is a mixture of code chunks/segments (sequences of commands in some programming language, e.g., R) and text chunks/segments (description of the problem, the code, and the results) and is written, formatted, and organized to be read by humans rather than a computer. As a result, this paradigm of writing software is unprecedentedly flexible in at least two ways: the source code can be i) extracted (*tangle*) from this document whenever needed (i.e., one tangles a document to get usable code); and ii) executed to get the output (*weave*; i.e., a program is woven to produce a document suitable for human viewing). The original intention of literate programming was to provide a mechanism for describing a program or algorithm. Although literate programming has never gained a large following, when coupled with other tools for testing and validating code, it provides a powerful mechanism for conveying descriptions, carrying out reproducible data analysis, and enhancing readability of the final document. This is the essential idea that dynamic document borrows from literate programming.

As the cornerstone of RR, dynamic document is similar to a computer program because it often contains the source code for all the analysis presented in the document. However, dynamic document is not a computer program, at least not a conventional computer program that is purely for computing. It also resembles a report or an academic paper whose mission is to document the study design, the data analysis, and more importantly interpret the results. Overall, a dynamic document consists of three essential parts: data, source codes, and narratives. A dynamic document is an ordered composition of code chunks and text chunks that describe and discuss a problem and its solution, and can be regarded as a source document from which the published static document can be generated. The content of the dynamic document is dynamically generated, e.g., figures, tables, and etc. are generated by executing the code chunks through *tangle*, and inserted into the document through *weave*. Thus, the contents including figures, tables, and etc. can be updated on the fly each time a view of the document is generated. Clearly, to obtain a dynamic document, we need a computing language for doing the analysis, and a documentation language for narratives.

To better illustrate the essence of a dynamic document, look at the following trivial example.

Suppose Mr. Traffic's job is to calculate and report traffic volume at a particular location (Location A). From 6 am to 7 am on the 6th of December 2010, Mr. Traffic obtained the following vehicle counts per 5 minutes through loop detectors: 115,95,125,110,78,118,113,88,97,106,105,101. Thus, in his report for that hour he wrote a single line: The volume at Location A from 6 am to 7 am on the 6th of December 2010 is 1251 vehicles per hour.

This is how a report is usually produced. Nothing is really wrong with that except that it has to be manually maintained and updated whenever there is any change due to its static nature. For example, later Mr. Traffic was told that two vehicle counts are wrong: 125 should be 120, and 78 should be 98. Since the way Mr. Traffic wrote his report is static, he would have to re-calculate the volume by adding these vehicle counts again, and then manually change the volume number from 1251 to 1266.

Of course, for this extremely simple example, it is not a big deal for Mr. Traffic to manually change the volume number (it would still be a big headache and error-prone if there are numerous numbers that need to be manually updated and are scattered throughout the whole document). However, this way of documenting and reporting is tedious and error prone. When the problem becomes more complex, this way of updating the document and report will not only be time-demanding, error-prone but also make the entire process much less transparent, and less reproducible. Even for Mr. Traffic himself, after a while, it would not be surprising if he couldn't remember how exactly he came up with these two different volumes and why and how he did so.

Now, image what if Mr. Traffic knows how to make his report dynamic by writing the one line report like this: The average volume at Location A from 6 am to 7 am on the 6th of December 2010 is {source code for computing the volume dynamically} vehicles per hour. One example using **R** is:

```
The volume at Location A from 6 am to 7 am on the 6th of December 2010 is r
sum(c(115,95,120,110,98,118,113,88,97,106,105,101)) vehicles per hour.
```

The above line produces: The volume at Location A from 6 am to 7 am on the 7th of July 2017 is 1266 vehicles per hour.

However, before we dive into technicalities of coding a dynamic document, let us take one step back to discuss some important high-level issues on how to plan a reproducible research from the scratch.

## A reproducible procedure for Transportation Research

Making a research project reproducible is a non-trivial task, and requires a careful planning, willingness to learn, and passion to share. This section presents a procedure for researchers, engineers and practitioners in Transportation Research in order to make their research reproducible at each stage.

A typical workflow of a research project in Transportation Research is illustrated in Figure 1. Overall this workflow consists of five stages: Study Design, Data Collection and Input, Data Processing, Data Analysis and Modelling, and Documentation and Dissemination. Among them, Stage II (Data Processing) is optional. State II and Stage III (Data Analysis and Modelling) are where most of the computational activities occur. Moreover, this workflow flows along an opposite direction between an author and a reader. The former starts with study design, goes through other stages, and ends with publications, while the latter starts with the final product: publications, and attempts to trace back to how the study was designed, how the data were collected, processed, and modeled, etc. In the current practice of academic publication, often the final product of a study (i.e., publications) is the only information source available to a reader. Solely relying on the publications a reader has to trace back how the study was designed, how the data were collected and processed, how the model was developed, etc. Obviously, it would be very difficult (if not impossible) for a reader to reproduce the analysis. To make a study truly reproducible, all the stages of the workflow need to be coded (using a computing language for the analysis, and a documentation language for narratives), and integrated as a dynamic document.

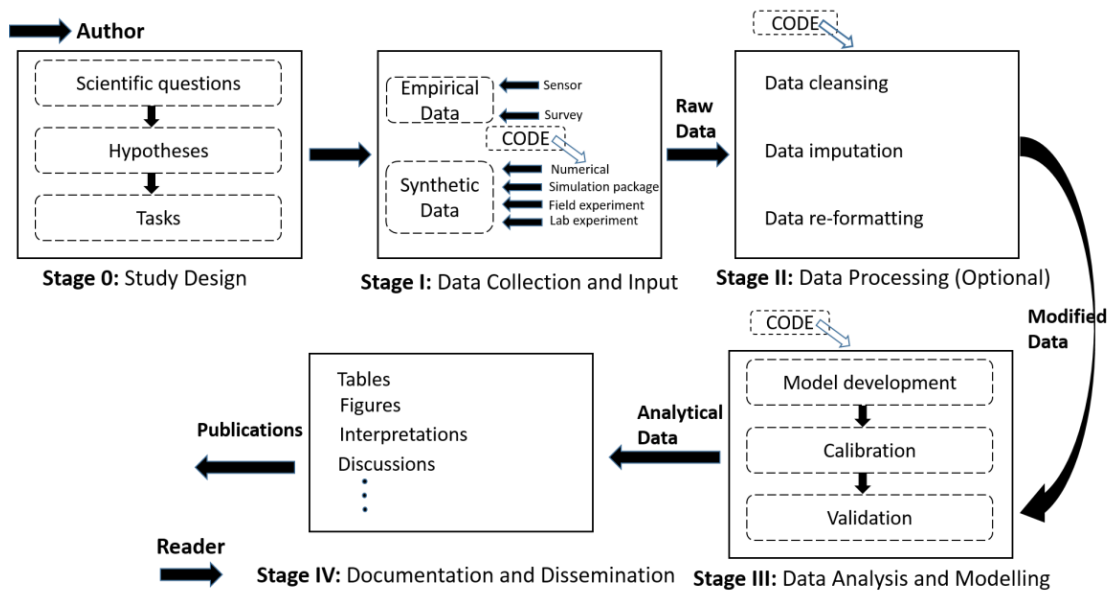

*Figure 1 A workflow of research project in Transportation Research*

## Tools for doing reproducible research

As briefly discussed above, over the past two decades, many tools for RR have been developed. Some are domain specific, and some are general. To the author's best knowledge, no domain-specific tools are available in Transportation Research. Thus, this paper focuses on general tools that can be adopted by researchers in Transportation Research to conduct RR. More specifically, this paper introduces tools based on **R** or compatible with **R**. Note that **RStudio**, an integrated development environment of **R** is used in the following discussion because **RStudio** is more user friendly with several features specifically designed for facilitating RR.

The first question is: why choose **R** for RR? Roughly speaking, a computing language suitable for conducting RR should be free and open-source, easy to use, and easy to share. **R** is a computing language specifically designed for statistical and other computational analysis. It is open-source and freely available for non-commercial use. Because of its huge user base, cutting-edge data analysis techniques can quickly become available in **R**. Its popularity has been consistently and rapidly rising among data analysists, and it has been consistently ranked as one of the top languages used for computational analysis (Gandrud 2016). More importantly, many widely used tools for conducting RR can be easily used in **R**. On the other hand, using commercial software like **MATLAB** violates the Freeware Principle (De Leeuw 2001), and inherently unfair to students. In addition, commercial software is closed, which means that its properties and credibilities have to be taken more or less on faith (Fateman 1992)).

Next, main tools for conducting RR at each step are introduced.

(1) Data storage: public folder in Dropbox, GitHub, or similar service providers.

In principle, data used for your analysis should be available and accessible to other researchers, as advocated by Open Data[1]. Numerous issues can arise regarding where and how data should be stored, e.g., security, version control, and etc. To solve these issues, free or low-cost cloud-based storage services are nice options. More specifically, for small size data files, they can be stored using GitHub, which is a popular platform primarily for software development.[2] File size policy of GitHub can be found at this link: https://help.github.com/articles/what-is-my-disk-quota/.

---

[1] https://en.wikipedia.org/wiki/Open_data

For large-size data files, they can be stored in Dropbox, Google Drive, or other similar service providers.

For huge data files, they can be segmented into smaller ones. For the purpose of demonstrating your work's reproducibility, it's unlikely the entire dataset is needed anyway.

Data stored in a GitHub account or other cloud-based platforms can be easily shared with other researchers, and imported into R.

(2) Data analysis: **R**, **Markdown**, **bookdown**

Computational analysis can be done by using **R**, and relevant code chunks are then embedded into the document through Markdown (In **RStudio**, a tool called **R Markdown** is available). More importantly, **R Markdown** allows you to embed **R** code chunks in your document to make the analysis transparent and reproducible, so that anyone (yes, including the future you) can trace back the entire journey of each part of your data analysis, will be able to know exactly how a particular model, table, figure, or a number has been obtained, and can easily recycle, modify, or extend your analysis. Needless to say, this is a huge benefit, not only to your readers, but also to yourself (just open a document authored by you two or three months ago, and ask yourself a simple question: Do you still remember how exactly you have conducted the analysis for this document? or open some of your recent codes to see whether you can quickly recall the meaning and purpose of each line).

There are many other benefits of using **R Markdown** as your data analysis workbench. **R Markdown** is well integrated with many other powerful, free, and open-source documenting tools, some are **R** packages such as **Knitr** (for this matter, any **R** package can be used in **R Markdown**), some are external, such as **LaTeX**. Within **R Markdown**, the final document can be generated in various output formats, e.g., Word document, PDF, html, and etc.

Recently, **RStudio** released one more powerful tool **bookdown** (developed by Yihui Xie, the same developer of the popular package **knitr**), which is built upon **Markdown** and **knitr**.

(3) Figures, tables and equations: **R**, **Markdown**, and **bookdown**

Figures, tables and equations are three typical elements of an academic document. To make a research reproducible, one of the technical challenges is to dynamically create figures and tables, and link them with the relevant narratives, which can be conveniently done in **RStudio** through **R Markdown** or **bookdown**. The way to include figures are essentially the same as for including other types of computational analysis. To create and include a table, you can either create a table from scratch, and embed the code chunk into the markdown file, or convert R objects to tables using table creation functions provided in several **R** packages, such as **knitr**, **xtable**, **texreg**, etc.

In addition, through **R Markdown** you have access to the powerful and flexible math mode of **LaTeX** to type, edit, and display various symbols and complicated mathematical expressions.

Examples on creating figures, tables, and equations in R Markdown are shown later in this paper.

(4) Documentation: **R**, **R Markdown**, and **bookdown**

The easiest way for documenting and interpreting any computational analysis in **R** is to use **R Markdown** or **bookdown**. Except the parts that special formatting is needed (most of the special formatting can be easily done using simple **Markdown** syntax), documenting using **R Markdown** is essentially the same as documenting using any text editor. A big advantage of using **R Markdown** is its capability of allowing you to conduct your data analysis right at the same place where you document and interpret your analysis, that is, you can directly code your data analysis using **R** in **R Markdown**, and mix your **R** code chunks with other parts of your document. By default, these **R** code chunks will be automatically executed and the outputs will be included in the final document when you generate the final report[3]. This powerful feature makes **R Markdown** a convenient platform for producing reproducible research. In **RStudio**, with a simple click **knitr** can be used to generate **R Markdown** files into various output formats (including html, PDF, word, etc.).

---

[2] https://github.com/

An **R Markdown** file essentially is a plain text file with a special extension, that is, *.Rmd*. Once you get used to the idea of using **R Markdown** as your reproducible and executable notebook, you will soon realize that writing a document in **R Markdown** is as convenient as writing a document in MS Word, but **R Markown** gives you 100% control on every detail of your document, and gives you options on how you would like to generate the final document.

A typical work flow of using **R Markdown** to document your data analysis procedure and produce a final report in various formats is outlined below:

First, you create an **R Markdown** file with an extension *.Rmd*, then the *.Rmd* file is knitted using **knitr**; as a result, a new **Markdown** file is created with an extension *.md*. The newly generated *.md* file is then processed by **pandoc**, and finally, the final report in various formats (e.g., html, MS Word, and PDF) is generated.

The workflow above may look a bit complicated, which involves a couple of packages you may not be familiar with. The good news is, in **RStudio** most of these steps are running silently without you even noticing it. You devote most of your time and effort to conducting and honing in your data analysis by writing the *.Rmd* file in the **R Markdown** interface provided by **RStudio**, as a data analyst should be doing, without being distracted by issues related to how to generate the final report. Once you are satisfied with your data analysis and the content of the report, generating a final report is literally as simple as clicking a button (the *knit* button).

In addition, you can reference, cross-reference, add footnote, and check spelling in **R Markdown**. Details can be found in Xie, Allaire, and Grolemund (2018).

This paper was entirely written by using **R Markdown**.

(5) Collaboration & feedback: Git and GitHub.

Version control is designed to manage source code by keeping track of changes to files. It can be naturally extended for RR because dynamic document, the cornerstone of RR can be regarded as a file of source code. Version control tools are needed to facilitate collaboration among researchers, disseminate research results to readers, and interact with readers by seeking feedback from readers and refining research results accordingly if needed.

Git and GitHub are popular verson control tools among software developers and programmers. Since writing a paper as a dynamic document is very similar to software development, Git and GitHub are naturally useful for RR, too. More specifically, Git is a useful tool for the author to make changes to and control versions of the dynamic document at the local computer; Git is directly integrated into **RStudio** projects. And GitHub is a useful platform for the purpose of collaborating with coauthors during the paper writing period and interacting with readers after the paper is published.

---

[3] You can manipulate how code chunks should be run and displayed to suit your specific purpose

Introducing how to use Git and GitHub is beyond the scope of this paper. Interested readers can refer to Chacon and Straub (2014).

## A complete example

In this section, a complete example on producing reproducible research in Transportation Research is given, based on a recent paper (Zheng and Washington 2012).

Note that since this is meant to be a "Hello, World!" example, to reduce its complexity and length, some of the techniques discussed in (Zheng and Washington 2012) are not covered in this example, such as second-order difference of cumulative data, short-time Fourier transform, etc. However, **this simple example includes all the typical features (e.g., headings, word formatting, references, inserting and cross-referencing figures, tables and equations, footnote, and etc.) an author of an academic paper may encounter when authoring a journal paper**, except data collection and storage (because numerical simulation was used). To save space, the example is not replicated below (because it is just a mini version of Zheng and Washington (2012) except that its entire content is coded as a dynamic document), but the source code of this example can be downloaded from http://www.connectedandautonomoustransport.com/uploads/2/5/2/6/25268286/reproducible_research_example.rmd.

## Challenges and a way for moving forward

This paper has introduced what RR is, why researchers in Transportation Research should make their computational analysis papers reproducible, and how RR can be done without much extra effort using tools freely available.

Although RR has been gaining lots of momentum and is at the verge of fundamentally changing how scientific work related to computational analysis should be conducted, documented, disseminated, and maintained, there exist many challenges of practicing RR. Some of these challenges are technical, some are cultural or behavioral (Peng 2011; Gandrud 2016).

One of the main technical challenges of practicing RR is the so called "dependency hell" (Guo 2013; Boettiger 2015). Operational systems and software are continuously changing for various reasons. Some changes are updates for better security or for better functionality, some changes are more dramatic, e.g., some version of the system becomes obsolete, and no longer supported or maintained. Because of these constant changes, a computational analysis that is original reproducible may become irreproducible later on. There is no easy solution for this problem. However, to alleviate this problem to some degree, an author can record the software environment as part of the dynamic document, which can be easily done in R by using the function `sessionInfo()`. Such information can clearly tell readers which version of the software is used, and what are the relevant packages, etc. The software environment of generating this paper can be found in Appendix.

Another main technical challenges of practicing RR is that researchers face significant barriers in learning these tools and approaches which are not part of their typical curriculum. In addition, currently there is a lack of incentives commensurate with the effort required for learning and practicing RR (FitzJohn et al. 2014; Joppa et al. 2013).

However, as pointed out by Carl Boettiger (Boettiger 2015), cultural and behavioral factors in many fields are a far more extensive primary barrier to reproducibility than these technical barriers, although lowering technical barriers can influence the cultural landscape as well. Caused by the power of habit, the primary barrier to computational reproducibility stems from a reluctance to publish the code used in generating the results (Boettiger 2015). Meanwhile, lack of requirements or incentive makes it more challenging to change researchers' habit. Although existing tools (some are introduced in this paper) makes RR accessible to any researcher, RR still requires extra time and conscientious effort from researchers, at least learning these tools needs time and effort (FitzJohn et al. 2014; Stodden 2010). Without requirement or incentive, it is hard to change the status quo because of the ever-increasing pressure in academia to publish, and to publish a lot, i.e., the "publish or perish" phenomenon.

Intuitively, funding agencies' policy, publishers' policy, and institutions' policy can play a critical role in changing the way researchers conduct research. As the primary venue for researchers to publish and share their research findings, more and more journals are embracing RR-oriented policies in their manuscript review process. *Biostatistics* introduced computational reproducibility. Journal *Computing in Science and Engineering* had a special issue on RR in early 2009, covering a variety of fields ranging from earth sciences to signal processing. (Stodden, Guo, and Ma 2013) conducted an empirical analysis of data and code policy adoption toward reproducible computational research by 170 journals in 2011 and 2012. 38% of these journals had a data policy, 22% had a code policy, and 66% had a supplemental materials policy as of June 2012. Many transportation related journals (including Transportation Research Part A-F) published by Elsevier encourage authors to share data, software, code, models, algorithms, protocols, etc. (https://www.elsevier.com/journals/transportation-research-part-b-methodological/0191-2615/guide-for-authors).

At individual level, the habit of doing things manually needs to be changed, particularly the habit of pointing and clicking cultivated from using Windows and other GUI programs (Sandve et al. 2013). Awareness on RR needs to be increased, and training needs to be provided for using existing tools to address the challenges of computational reproducibility.

Finally, besides the dynamic document approach introduced in this paper, there are other approaches for doing RR, including workflow solutions (Dudley and Butte 2010), virtual machines approach (Howe 2012), DevOps (Development and Systems Operation) approach (Boettiger 2015). Also, RR is not the same as independent verification, and does not necessarily verify the conclusions or inferences about the subject matter, although verifying a reproducible study is generally easier.

Note that this paper has been written entirely in a reproducible way using the tools introduced above. The source code of the corresponding dynamic document can be downloaded from GitHub: xxx (the link will be added once the paper is accepted for publication) or from the link: http://www.connectedandautonomoustransport.com/xxx (the link will be added once the paper is accepted for publication). Instructions on how to generate the exact paper that you are reading and the software environment in which this paper has been written can also be downloaded from the same link. Moreover, the dynamic document of this paper can be treated as a template for transportation researchers who want to prepare their next journal paper in a reproducible way.

**Funding Statement**: This research is partially funded by the Australian Research Council (ARC) through Dr. Zuduo Zheng's Discovery Early Career Researcher Award (DECRA; DE160100449). The funder was not involved in the manuscript writing, editing, approval, or decision to publish.

**The Conflict of Interest Disclosure**: The author(s) declare(s) that there is no conflict of interest regarding the publication of this paper.

**Data Availability**: This article was written entirely in a reproducible way using **Rstudio** and **RMarkdown** from head to toe. **The source code for generating the exact document you are reading was submitted to the journal as supplementary material for the review purpose**, and can be downloaded from GitHub: xxx (the link will be added once the paper is accepted for publication) or from the webpage: http://www.connectedandautonomoustransport.com/xxx (the link will be updated once the paper is accepted for publication).

# Appendix

The software environment of generating this paper is shown below:

```
sessionInfo()

## R version 3.4.3 (2017-11-30)
## Platform: x86_64-w64-mingw32/x64 (64-bit)
## Running under: Windows 10 x64 (build 14393)
##
## Matrix products: default
##
## locale:
## [1] LC_COLLATE=English_Australia.1252  LC_CTYPE=English_Australia.1252
## [3] LC_MONETARY=English_Australia.1252 LC_NUMERIC=C
## [5] LC_TIME=English_Australia.1252
##
## attached base packages:
## [1] stats     graphics  grDevices utils     datasets  methods   base
##
## other attached packages:
## [1] TTR_0.23-3
##
## loaded via a namespace (and not attached):
##  [1] Rcpp_0.12.15    bookdown_0.6    lattice_0.20-35 zoo_1.8-1
##  [5] digest_0.6.14   rprojroot_1.3-2 grid_3.4.3      backports_1.1.2
##  [9] magrittr_1.5    evaluate_0.10.1 highr_0.6       stringi_1.1.7
## [13] curl_3.1        xts_0.10-1      rmarkdown_1.8   tools_3.4.3
## [17] stringr_1.3.1   xfun_0.5        yaml_2.1.16     compiler_3.4.3
## [21] htmltools_0.3.6 knitr_1.18
```